\documentclass[12pt,preprint]{aastex}

\slugcomment{Version: \today}

\def\deg{\hbox{$^\circ$}}

\shorttitle{Spectral Evolution of NGC\,1275 with {\it Fermi} LAT}
\shortauthors{Kataoka et al.}  
\usepackage{times}

\begin{document}

\title{$\gamma$-ray Spectral Evolution of NGC\,1275 Observed with {\it Fermi} LAT}

\author{J. Kataoka$^{1}$, \L . Stawarz$^{2,\,3}$, 
C.C. Cheung$^{4}$, G. Tosti$^{5,\,6}$, E. Cavazzuti$^{7}$, A. Celotti$^{8}$,
S. Nishino$^{9}$, Y. Fukazawa$^{9}$, D.~J.~Thompson\altaffilmark{10}, W. F. McConville\altaffilmark{10,11}}

\altaffiltext{1}{Research Institute for Science and Engineering, Waseda University, 3-4-1, Okubo, Shinjuku, Tokyo, 169-8555, Japan}
\altaffiltext{2}{Institute of Space and Astronautical Science, JAXA, 3-1-1, Yoshinodai, Sagamihara, Kanagawa, 252-5210, Japan} 
\altaffiltext{3}{Astronomical Observatory, Jagiellonian University, ul. Orla 171, Krak\'ow 30-244, Poland}
\altaffiltext{4}{NRC Research Associate, Space Science Division, Naval Research Laboratory, Washington, DC 20375, USA}
\altaffiltext{5}{Istituto Nazionale di Fisica Nucleare, Sezione di Perugia, I-06123 Perugia, Italy}
\altaffiltext{6}{Dipartimento di Fisica, Universit\`a degli Studi di Perugia, I-06123 Perugia, Italy}
\altaffiltext{7}{Agenzia Spaziale Italiana (ASI) Science Data Center, I-00044 Frascati (Roma), Italy}
\altaffiltext{8}{Scuola Internazionale Superiore di Studi Avanzati (SISSA), 34014 Trieste, Italy}
\altaffiltext{9}{Department of Physical Sciences, Hiroshima University, Higashi-Hiroshima, Hiroshima 739-8526, Japan}
\altaffiltext{10}{NASA Goddard Space Flight Center, Greenbelt, MD 20771, USA}
\altaffiltext{11}{Department of Physics, University of Maryland, College Park, MD 20742, USA}

\begin{abstract}

We report on a detailed investigation of the high-energy $\gamma$-ray 
emission from NGC\,1275, a well-known radio galaxy hosted by a giant 
elliptical located at the center of the nearby Perseus cluster. 
With the increased photon statistics, the center of the $\gamma$-ray 
emitting region is now measured to be separated by only $0.46$\,arcmin 
from the nucleus of NGC\,1275, well within the $95\%$ confidence 
error circle with radius $\simeq 1.5$\,arcmin.
Early {\it Fermi}-LAT observations revealed a significant decade-timescale 
brightening of NGC\,1275 at GeV photon energies, with a flux about 
seven times higher than the one implied by the upper limit from 
previous EGRET observations.  
With the accumulation of one-year of {\it Fermi}-LAT all-sky-survey 
exposure, we now detect flux and spectral 
variations of this source on month timescales, as reported in this 
paper. The average $>$100 MeV $\gamma$-ray spectrum of NGC\,1275 shows 
a possible deviation from a simple power-law shape, indicating a 
spectral cut-off around an observed photon energy of 
$\varepsilon_{\gamma} = 42.2 \pm 19.6$ GeV, with an average flux of 
$F_{\gamma} = (2.31 \pm 0.13) \times 10^{-7}$\,ph\,cm$^{-2}$\,s$^{-1}$ 
and a power-law photon index, $\Gamma_{\gamma} = 2.13 \pm 0.02$. 
The largest $\gamma$-ray flaring event was observed in 
April--May 2009 and was accompanied by significant 
spectral variability above $\varepsilon_{\gamma} \gtrsim 1-2$ GeV. The 
$\gamma$-ray activity of NGC\,1275 during this flare can be described 
by a hysteresis behavior in the flux versus photon index plane. The 
highest energy photon associated with the $\gamma$-ray source was detected 
at the very end of the observation, with the observed energy of 
$\varepsilon_{\gamma} = 67.4$\,GeV and an angular separation of about 
$2.4$\,arcmin from the nucleus. In this paper we present the details 
of the {\it Fermi}-LAT data analysis, and briefly discuss the 
implications of the observed $\gamma$-ray spectral evolution of 
NGC\,1275 in the context of $\gamma$-ray blazar sources in 
general.

\end{abstract}


\section{Introduction}
\label{sec:intro}
With the successful launch of the {\it Fermi} 
Gamma-ray Space Telescope, 
we have a new opportunity to study the
$\gamma$-ray emission from different types of extragalactic sources ---
not only blazars, but also radio galaxies and possibly other classes of
active galactic nuclei (AGN) --- with much improved
sensitivity than previously available (Abdo et al. 2010a). 
During the initial all-sky
survey performed during the first four months after its launch, the {\it
Fermi} Large Area Telescope (LAT) detected only two radio galaxies at high significance ($\geq
10\sigma$), namely NGC\,1275 (Abdo et al. 2009a; hereafter Paper-I) and
Cen\,A (Abdo et al. 2009b, 2009c). More recently, the detection of MeV/GeV emission from
yet another famous radio galaxy M\,87, an established TeV source, 
was reported based on ten-months of all-sky survey {\it Fermi}-LAT data
(Abdo et al. 2009d). Yet the detection of NGC\,1275 was particularly
noteworthy  because this source, unlike Cen\,A or M\,87, was previously 
undetected in $\gamma$-rays, neither by {\it CGRO}/EGRET
during its $\sim$10 years of operation, nor by ground-based Cherenkov
Telescopes. The $\gamma$-ray flux of NGC\,1275 detected by the {\it
Fermi}-LAT was about seven times higher than the one implied by the
$2\sigma$ upper limit reported by EGRET, namely $F_{\varepsilon_{\gamma}
> 100\,{\rm MeV}} < 3.72 \times 10^{-8}$\,ph\,cm$^{-2}$\,s$^{-1}$
(Reimer et al. 2003). We note that {\it COS B} data taken between
1975 and 1979 (Strong et al. 1982; Mayer-Hasselwander et al. 1982)
showed a $\gamma$-ray excess coincident with the position of this galaxy, although
the evidence for the claimed high-energy source to be \emph{uniquely}
related to NGC\,1275 is ambiguous.  

NGC\,1275 is a giant elliptical
galaxy located at the center of the Perseus cluster\footnote{The Perseus
cluster (Abell 426): redshift $z=0.0179$, luminosity distance $d_{\rm L} =
75.3$\, Mpc, scale $21.5$\,kpc arcmin$^{-1}$ (for flat cosmology with $H_{\rm 0} =
71$\,km\,s$^{-1}$\,Mpc$^{-1}$ and $\Omega_{\rm M} = 0.27$).}. This cluster is
the brightest cluster of galaxies in the X-ray band (e.g.,
B\"{o}hringer et al. 1993; Fabian et al. 2003, 2006), and as such it has
been the focus of several extensive research programs over many years
and across the entire available electromagnetic spectrum. When observed at
radio wavelengths, NGC\,1275 hosts the exceptionally bright radio FR\,I radio galaxy
Perseus\,A = 3C\,84 (e.g., Vermeulen
et al. 1994, Taylor et al. 1996, Walker et al. 2000, Asada et
al. 2006). Although high-energy $\gamma$-rays may in general be produced
within the intergalactic/interstellar medium of the Perseus cluster, in
Paper-I, we argued that the inner radio jet of 3C\,84 was
the most likely source of the observed $\gamma$-ray photons 
because of the variability of the MeV/GeV flux on year/decade timescales
implied by the EGRET and early {\it Fermi}-LAT observations. 
Specifically, these
measurements implied the $\gamma$-ray emission region size in NGC\,1275 has a 
radius, $R \lesssim ct_{\rm var} \simeq 1$\,pc. Note, however, that no significant variability was detected within the
four-month long {\it Fermi}-LAT dataset analyzed in Paper-I.
Also, the LAT error circle determined
from these initial data was too large to exclude possible
contributions from other galaxies to the observed $\gamma$-ray flux
(specifically, NGC~1273, 1274, 1277, 1278 and 1279, were still
within the previously determined $95\%$ error circle with $R=5.2$\,arcmin).

Meanwhile, follow-up observations by the VERITAS Cherenkov telescope
(Weekes et al. 2002) put strong constraints on the very high-energy
(VHE) $\gamma$-ray emission from NGC\,1275 above 100 GeV. In particular,
no VHE $\gamma$-ray emission from NGC\,1275 was detected by VERITAS,
with a $99\%$ confidence level upper limit of $2.1\%$ of the Crab Nebula
flux, corresponding to $19\%$ of the power-law extrapolation of the MeV/GeV
flux observed during the first four months of the {\it Fermi}-LAT
observations (assuming the photon index, $ \Gamma_{\gamma} \simeq 2.2$;
Acciari et al. 2009b). This naturally indicates a deviation from the
pure power-law spectrum in the VHE regime, possibly compatible with the
presence of an exponential cutoff around or below photon energies
$\varepsilon_{\gamma} \simeq 100$\,GeV (Acciari et al. 2009b). 
The MAGIC Cherenkov telescope also recently measured
upper limits for the VHE $\gamma$-ray emission of NGC\,1275, namely
$F_{\varepsilon_{\gamma} > 100\,{\rm GeV}} < (4.6-7.5) \times
10^{-12}$\,ph\,cm$^{-2}$\,s$^{-1}$ for the photon indices ranging from
$1.5$ up to $2.5$ (Aleksi\'c et al. 2010). Thus, the implied deviation in
the $\gamma$-ray spectrum of 3C\,84 from a simple power-law form, as
well as a possibility for a short-timescale ($<$ month) variability of
the Perseus\,A $\gamma$-ray flux, may now be finally addressed and
re-examined by {\it Fermi}-LAT, due to the much improved photon
statistic (especially above $10$\,GeV) after the one-year-all-sky
survey. Obviously, such deep studies of NGC\,1275 in the MeV/GeV photon
energy range are of the major importance for understanding the whole
class of $\gamma$-ray emitting radio galaxies in general. 

Firmly motivated, we performed a detailed investigation of
NGC\,1275 in $\gamma$-rays based on the accumulation of one-year of {\it
Fermi}-LAT all-sky survey data. In particular, we aimed
to address the following problems: (i) presence of short-timescale
$\gamma$-ray flux variability, (ii) positional coincidence of the
$\gamma$-ray emitting center with the active nucleus of NGC\,1275, and
(iii) spectral curvature and spectral evolution of NGC\,1275 in the MeV/GeV
photon energy range. In $\S$\,2, we describe the {\it Fermi}-LAT
$\gamma$-ray observations and data reduction procedure. The results of the
analysis are given in $\S$\,3, and the discussion and conclusions are presented
in $\S$\,4.

\section{{\it Fermi}-LAT Observations} 
\label{sec:observations}
The LAT instrument onboard {\it Fermi} is described in 
detail in Atwood et al. (2009) and references
therein. Compared to earlier $\gamma$-ray missions, the LAT has a large
effective area ($\sim$$8,000$\,cm$^2$ on axis at $1$\,GeV for the event
class considered here), wide energy coverage (from $\approx 20$\,MeV to
$>300$\,GeV), and improved angular resolution. The $68\%$
containment angles of the reconstructed incoming photon direction are
approximated as $\theta_{68}\simeq
0^{\circ}.8\,(\varepsilon_{\gamma}/{\rm GeV})^{-0.8}$ below $10$\,GeV,
giving $\theta_{68} \sim 5^{\circ}.1$ at $100$\,MeV and $\sim
0^{\circ}.15$ at $10$\,GeV (update of values presented in Atwood et
al. 2009). Above $10$\,GeV, the improvement of angular resolution
becomes relatively gentle, such that $\theta_{68} \sim 0^{\circ}.07$ 
at $100$\,GeV.
During the first year of operations, most of the
telescope's time was dedicated to observing in ``survey mode'' 
where {\it Fermi} points away from the Earth and nominally rocks the
spacecraft axis north and south from the orbital plane to enable
monitoring of the entire sky on a time scale shorter than a day. 
The whole sky is surveyed every $\sim 3$\,hours (or 2 orbits). 
The total live time included is 280.3 days (24.21 Ms). This corresponds 
to an absolute efficiency of 75.0 $\%$. Most of the inefficiency is due 
to time lost during passages through the South Atlantic Anomaly and 
to readout dead time.


The observations used here
comprises all scientific data obtained between 4 August 2008 and 13
August 2009. This time interval runs from Mission Elapsed Time (MET) 
239557417 to 271844550. We have
applied the zenith angle cut of $105^{\circ}$ to eliminate photons from the Earth's limb.
The same zenith cut is also accounted for in the exposure
calculation using the LAT science tool\footnote{http://fermi.gsfc.nasa.gov/ssc/data/analysis/documentation/Cicerone/}
\textsc{gtltcube}. We use the
``Diffuse'' class events (Atwood et al.\ 2009), which are those
reconstructed events having the highest probability of being photons. In
the analysis presented here, we set the lower energy bound to a value of
$100$\,MeV. 
Science Tools version v9r15p2 and IRFs (Instrumental Response Functions) \textsc{P6\_V3} (a model of
the spatial distribution of photon events calibrated post-launch) were
used throughout this paper. The early portion of the data here 
coincides with the early LAT observations of NGC\,1275 presented in Paper-I (4 August 2008
and 5 December 2008).

\subsection{Results}
\label{sec:results}

Figure\,1 shows the LAT $95\%$ $\gamma$-ray localization 
error circles placed on the {\it Chandra} ACIS-S X-ray 
image of NGC\,1275/Perseus region (ObsID 4952;
exposure $164$\,ks). The {\it Chandra} image is constructed between
$0.4$ and $8.0$\,keV. 
The positional center of the $\gamma$-ray emission (RA$= 49\deg.941$, DEC$=
41\deg.509$) is taken from the 11-month catalog (1FGL~J0319.7+4130; Abdo et
al. 2010b), and is only $0.46$\,arcmin from the NGC\,1275 nucleus 
(RA$ = 49\deg.951$, DEC$ = 41\deg.512$), with $95\%$ radius 
$\simeq 1.5$\,arcmin (Fig.~1). Since the localization
error has been substantially improved compared to the one quoted in Paper-I
($5.2$\,arcmin, based on the 3-month bright source list; Abdo et al.2009c),
all the other field galaxies mentioned in Paper~I within the 3-month LAT
error (NGC~1273, 1274, 1277, 1278, and 1279) are now outside the 11-month
LAT error circle. Indeed, even the nearest galaxy, NGC~1274, is 2.6 arcmin
apart from the positional center of the $\gamma$-ray emission with 95 $\%$
radius of 1.5 arcmin. With the good positional association
described above, we thus believe the $\gamma$-ray source is most 
likely identified as the nucleus of NGC~1275, but still needing further 
confirmation based on correlated variability with observations at 
other wavelengths.

We also checked the projection of the $\gamma$-ray images as presented in
Paper-I. The counts distributions of NGC\,1275 in different energy 
bands are consistent with the
distributions for a point source, indicating that the diffuse extended
component does not contaminate the NGC\,1275/Perseus $\gamma$-ray source,
even with much improved photon statistics in the one-year 
dataset. Since the angular resolution of the LAT improves at high
energies (Atwood et al. 2009 and see above), it is interesting to
compare the reconstructed incoming photon directions with the position
of the NGC\,1275 nucleus around the higher energy photons that can be
detected with the {\it Fermi}-LAT. Figure\,2 shows the angular
separation of $\varepsilon_{\gamma} > 10$\,GeV photons from the nucleus
of NGC\,1275 as a function of photon energy. The $68\%$ and $95\%$
angular resolutions of {\it Fermi}-LAT are shown as dotted and dashed
lines, respectively. The angular displacement of the highest energy
photon detected during one-year-all-sky survey ($\varepsilon_{\gamma}
\simeq 67.4$\,GeV; a double circle with a number ``I'' in Figure\,2)
from the nucleus is only $2.4$\,arcmin, which is well within the PSF of
the LAT at this energy ($\theta_{68} \simeq 4.4$\,arcmin). Note that the
chance probability for detecting Galactic and/or extragalactic
$\gamma$-ray background photons with energies 
at least $67.4$\,GeV within $\theta_{68}$ is 
less than $0.1\%$, hence we conclude that NGC\,1275 is the
most likely source of the discussed photon. The other three highest
energy photons ($\varepsilon_{\gamma} > 30$\,GeV) are marked in
Figure\,2 as ``II'', ``III'' and ``IV'', and are all well
within the $95\%$ PSF of the {\it Fermi}-LAT. 

\subsection{Spectral Analysis}
\label{sec:spectra}

To study the average spectrum of NGC\,1275 during the one-year 
observation, we use the standard maximum-likelihood spectral estimator 
provided with the LAT science tools \textsc{gtlike}.  This fits the 
data to a source model, along with the models for the uniform 
extragalactic and structured Galactic backgrounds\footnote{http://fermi.gsfc.nasa.gov/ssc/data/access/lat/BackgroundModels.html}. We use a recent 
Galactic diffuse model, gll\_iem\_v02.fit, with the normalization 
free to vary in the fit. The response function used is 
\textsc{P6\_V3\_DIFFUSE}. Careful choice of the source region is 
important especially for relatively faint sources. Following the 
detailed study of changing the region of interest (ROI) radius from 
$5^{\circ}$ to $20^{\circ}$ in Paper-I, we set $r = 8^{\circ}$ in the 
following analysis to minimize the contamination from the Galactic 
diffuse emission. Only one point source, 
corresponding to src\_A in Paper-I, was found in the ROI.
The coordinates are RA$ = 55\deg.105$, DEC$ = 41\deg.515$,  
with $r_{95} = 4.1$\,arcmin. We modeled the source for subsequent 
spectral analysis (also see the {\it Fermi}-LAT 
bright $\gamma$-ray source list; Abdo et al. 2009c). 

We first model the continuum $\gamma$-ray 
emission of NGC\,1275 with a single power-law. The extragalactic 
background is assumed to have a power-law spectrum as well, with the 
spectral index and the normalization free to vary in the fit. From an 
unbinned \textsc{gtlike} analysis, the best fit power-law parameters for 
NGC\,1275 are:

\begin{equation}\frac{dN}{d\varepsilon_{\gamma}} = (2.61 \pm 0.14) 
\times 10^{-9} \;(\frac{\varepsilon_{\gamma}}{100\,{\rm 
MeV}})^{-2.13\pm0.02} \hspace{5mm}{\rm ph\,cm^{-2}\,s^{-1}\,MeV^{-1}} 
\, ,\end{equation}

\noindent or

\begin{equation}F_{\varepsilon_{\gamma}>100\,{\rm MeV}} = (2.31 \pm 
0.13) \times 10^{-7} \hspace{5mm}{\rm ph\,cm^{-2}\,s^{-1}} \, , 
\end{equation}

\noindent with only statistical errors taken into account. 
Systematic errors for the LAT are still under investigation 
,  but the estimated systematic uncertainty on the flux is 
10 $\%$ at 100 MeV, 5 $\%$ at 500 MeV, 
and 20 $\%$ at 10 GeV, respectively (Abdo et al. 2010b).
The results are consistent with those presented in Paper-I, even though the 
uncertainties are now much smaller due to the improved photon 
statistics.

The predicted photon counts from NGC\,1275 in the ROI are $N_{\rm pred}
= 3187.0$ and the test statistic (defined as $TS = 2\,[\log L - \log
L_0$], where $L$ and $L_0$ are the likelihood when the source is
included or not, respectively) is $TS = 4039.8$ above
$\varepsilon_{\gamma} = 100$\,MeV, corresponding to a $64\sigma$
detection. For the Galactic diffuse background, the normalization is
$1.11 \pm 0.01$ and $N_{\rm pred} = 44835.3$. The power-law photon index
of the extragalactic background is $\Gamma_{\gamma} = 2.31 \pm 0.03$
with $N_{\rm pred} = 9081.7$.  Figure\,3 shows the LAT spectrum of
NGC\,1275 obtained by separately running \textsc{gtlike} for 10 energy
bands; $100-200$\,MeV, $200-400$\,MeV, $400-800$\,MeV,
$800$\,MeV\,$-1.6$\,GeV, $1.6-3.2$\,GeV, $3.2-6.4$\,GeV,
$6.4-12.8$\,GeV, $12.8-25.6$\,GeV, $25.6-51.2$\,GeV, and
$51.2-102.4$\,GeV, where the dashed line shows the best fit power-law
function for the NGC\,1275 data given in equation (1). 
For the highest energy bin 
($51.2-102.4$\,GeV), we plot a 2 $\sigma$ upper limit since this 
bin includes only one photon (see Figure. 2) and is ignored  
in the subsequent statistical analysis. 

Note an indication of a deviation of the model with respect to the data above
$\varepsilon_{\gamma} = 20$\, GeV. 
Indeed, a $\chi^2$ fit of the power-law model to the 
data\footnote{Note that the maximum
likelihood itself does not provide any information about the quality of
a fit for an assumed model. We therefore perform $\chi^2$ fitting to the
resultant 10 band LAT spectrum as described above, to give a convenient
estimate of the goodness of the fit.} gives a
relatively poor fit with $\chi^2 = 14.0$ for $7$ degrees of freedom
(d.o.f.), where its probability is $P(\chi^2) = 0.05$. 
Instead, an alternative fit with a
cutoff power-law function with the form $dN/d\varepsilon_{\gamma}
\propto \varepsilon_{\gamma}^{-\Gamma_{\gamma}} \times
\exp\left[-\varepsilon_{\gamma}/\varepsilon_{\rm c}\right]$ gives a much
better representation of the data with an improved $\chi^2 = 6.7$ for
$7$\,d.o.f., corresponding to $P(\chi^2) = 0.34$. We therefore retried the
\textsc{gtlike} fit assuming a power-law with an exponential cutoff
function, and obtained $\varepsilon_{\rm c} = 42.2 \pm 19.6$\,GeV 
with $\Gamma_{\gamma} = 2.07 \pm 0.03$.
Moreover, we apply a likelihood ratio test between a simple power-law and 
a cutoff power-law function. The test statistic ($D$) 
is twice the difference in these log-likelihoods, which 
gives $D$ = 2.9 for our case. Note that the probability distribution 
of the test statistic can be approximated by a $\chi^2$ distribution 
with 1 d.o.f., corresponding to different 
d.o.f. between two functions. We obtain  $P(\chi^2)$ $\le$ 0.08, 
which again indicates a deviation from a 
simple power-law function although this is currently inconclusive. The best-fit cutoff power-law function 
is shown as dotted line in Figure\,3. 

\subsection{Temporal Variability}
\label{sec:spectra}

Next, we investigated the $\gamma$-ray flux variations of
NGC\,1275 from 4 August 2008 to 13 August 2009. To this end, we
constructed spectra with a time resolution of 14 days and fit each
spectrum with a power-law model just for simplicity. 
The ROI radius ($r = 8^{\circ}$),
the energy range ($\varepsilon_{\gamma} > 100$\,MeV), and other
screening conditions are the same as described above. Since no
variability is expected for the underlying background diffuse emission,
we fix the best-fit parameters to the average values determined from
the one-year integrated spectrum for the Galactic and extragalactic
background components. We first fix the spectral index to the best fit
value over the full interval, $\Gamma_{\gamma} = 2.13$, to minimize
uncertainties in the flux estimates. In this case, variability is highly
significant with $\chi^2 = 157.7$ for $26$\,d.o.f., where $P(\chi^2) <
10^{-6}$. Indeed, the $\gamma$-ray flux varies even within a few month
timescale. Due to the limited photon statistics, however, it is difficult to 
investigate any shorter timescale variability of NGC\,1275. We also
investigate the spectral evolution of NGC\,1275, with the $\gamma$-ray
photon index free to vary. Figure\,4 shows variations of the flux
($\varepsilon_{\gamma} > 100$\,MeV: {\it upper panel}) and photon index
({\it lower panel}) versus time. A flaring event is seen around $T =
252-294$\,days after 2008 August 4th (epoch ``B'' in Figure\,4),
corresponding to the epoch April--May 2009. A doubling timescale 
for the flaring period cannot be accurately measured, but $\sim$20
days. 
A $\chi^2$ fit to a constant gives $35.6$ and $43.0$ 
for $26$\,d.o.f., corresponding to the
probability $P(\chi^2) = 0.10$ and $0.02$, respectively, for the flux
and the photon index variations. Interestingly, the spectral variation
is predominantly due to the difference between the pre- and post-flare
period, denoted in Figure\,4 as epochs ``A'' and ``C''. In fact, if we
test the variability \emph{separately} for the pre- and post-flare
periods (A and C), the significance of the variability decreases
substantially, such that a constant fit provides $P(\chi^2) > 0.21$ or
$0.29$ for both the flux and the photon index changes. During the flare
(epoch B), the photon index changes from $\Gamma_{\gamma} \simeq
2.2$ to $\Gamma_{\gamma} \le 2.0$. Note that, after the flare (epoch
C), the flux drops to its original level but the spectrum remains
relatively flat, with $\Gamma_{\gamma} \simeq 2$, persisting for more
than a few months. It should also be noted that three out of the 
four highest
energy photons (denoted as I, II, and III in Figure\,2) were indeed 
detected in
the post-flare period, when the spectrum appears the flattest.

\section{Discussion and Conclusion}
\label{sec:discussion}

In the previous sections, we reported on the analysis of the $\gamma$-ray
emission from NGC\,1275 observed with {\it Fermi}-LAT during its
one-year-all-sky survey. We showed that with the increased photon
statistics, the positional center of the $\gamma$-ray emission is 
now much closer to the NGC\,1275 nucleus as compared to that reported in
Paper-I. 
In addition, we have shown
that the average $\gamma$-ray spectrum of NGC\,1275 reveals a
significant deviation from a simple power-law above photon energies
$\varepsilon_{\gamma} \sim 1-2$\,GeV. That is, the observed {\it Fermi}-LAT 
spectrum is best fitted by a power-law function 
($\Gamma_{\gamma} \simeq$ 2.1) with an exponential cutoff at the
break photon energy $\varepsilon_{\rm c} = 42.2 \pm 19.6$\,GeV. Finally,
we argued that significant flux and spectral changes of NGC\,1275
are detected with {\it Fermi}-LAT on a timescale of a few months,
although the possibility for even shorter variability
remains uncertain. 

We also reported the detection of an interesting spectral evolution, 
consisting of a \emph{persistent} (over more than a few months)
spectral hardening (from $\Gamma_{\gamma} \simeq 2.2$ to
$\Gamma_{\gamma} \simeq 2$) after the largest flaring event observed in
April--May 2009. During this flat-spectrum/low-flux-level epoch the
highest energy photon ($\varepsilon_{\gamma} \simeq 67.4$\,GeV) was 
detected from the direction of NGC\,1275. All these new findings
basically support the idea put forward in Paper-I that the observed
$\gamma$-ray emission from the Perseus system originates in the (sub)
pc-scale radio jet of NGC\,1275, and is therefore most likely
analogous to high-energy emission observed in blazars. In fact, as shown in
Paper-I, the overall $\nu F_\nu$ spectral energy distribution (SED) of
NGC\,1275 constructed with multi-frequency radio to $\gamma$-ray data
shows a close similarity to the ``two-bump'' SEDs of so-called
low-frequency peaked BL Lac objects (hereafter LBLs). Till now, only a few 
LBLs have been detected at TeV photon energies -- BL Lac (Albert et
al. 2007),  3C\,66A (Acciari et al. 2009a), S5 0716+714 (Anderhub et
al. 2009), and also W~Comae from an IBL class (Acciari et al. 2008;
2009d)  -- and more
similar discoveries are expected. Hence, NGC\,1275 itself has
been suggested to be a potential TeV source as well, thus motivating deep
VERITAS (Acciari et al. 2009b) and MAGIC (Aleksi\'c et
al. 2010) observations. These observations so far resulted only in upper limits
to the VHE $\gamma$-ray emission of the studied region.
One should note, however, that both the low- and high-energy
peaks of NGC\,1275 (in the $\nu F_\nu$ representations) are located at
substantially lower frequencies than those of typical LBLs. Indeed, the
low-energy (synchrotron) emission components of 3C\,66A and BL Lac peak
around $10^{13-15}$\,Hz, while it is around
$10^{12}$\,Hz in the case of NGC\,1275 (Paper-I). Correspondingly, the
high-energy emission component of 3C\,66A (but not necessarily
of BL Lac) peaks at higher photon energies than that observed in
NGC\,1275. Within such an interpretation, this is consistent with
the softer MeV/GeV spectrum in NGC\,1275 ($\Gamma_{\gamma} = 2.13 \pm
0.02$) compared to 3C\,66A ($\Gamma_{\gamma}
= 1.97 \pm 0.04$; Abdo et al. 2009b) as measured by the {\it Fermi}-LAT.
Note also that the X-ray spectrum
of NGC\,1275 is clearly ``rising'' in the $\nu F_\nu$ representation
($\Gamma_{\rm X}$ = 1.6$-$1.7; Balmaverde et al. 2006, Ajello et al. 2009), 
indicating that it is dominated by the low-energy portion of the IC
emission, whilst the X-ray spectra of LBLs are typically very flat, 
suggesting a transition between
the synchrotron and IC components ($\Gamma_{\rm X}$ $\simeq$ 2; 
e.g., Ghisellini et al. 1998). Such
observational differences may indicate that, unlike in the case of LBLs, the
high-energy spectrum of NGC\,1275 does not extend up to TeV photon
energies.

An interesting comparison can be made to another nearby radio galaxy detected by {\it Fermi}-LAT, namely
M\,87 which resembles NGC\,1275 in many respects, and is an established TeV
source (Abdo et al. 2009d). While M\,87 is closer to us
than NGC\,1275 ($d_{\rm L} = 16$\,Mpc versus $d_{\rm L} =
75.3$\,Mpc), the GeV flux of M\,87 is much lower than that
of NGC\,1275 ($F_{\varepsilon_{\gamma}>100\,{\rm MeV}}
= (2.45 \pm 0.63) \times 10^{-8}$\,ph\,cm$^{-2}$\,s$^{-1}$ versus
$(2.31 \pm 0.13) \times
10^{-7}$\,ph\,cm$^{-2}$\,s$^{-1}$), with similar LAT measured spectra 
($\Gamma_{\gamma} = 2.26 \pm 0.13$ versus
$2.13 \pm 0.02$). Note also that both radio galaxies
are located at the centers of rich clusters, that the synchrotron
emission components in both sources peak in the far infrared ($\sim
10^{12}-10^{13}$\,Hz), and the estimated jet powers are similar, 
$L_{\rm j}\sim 10^{44}$ erg s$^{-1}$ 
(Owen et al. 2002, Dunn \& Fabian 2004). 
Thus, it may be surprising that only one of these 
has so far been detected at TeV photon energies. The detailed
analysis of the spectral evolution of NGC\,1275 within the {\it Fermi}-LAT 
range reported in this paper may provide a viable explanation for
such a behavior. In particular, as already emphasized above, we found
that the epochs characterized by the flattest GeV continuum of this
source, as well as the arrival times of the highest energy photons from
the direction of NGC\,1275, do not coincide with the epochs of the
highest photon flux above $100$\,MeV. This is clearly illustrated in
Figure\,5, which shows a correlation between
$F_{\varepsilon_{\gamma}>100\,{\rm MeV}}$ and the photon indices
emerging from the power-law fits. 
Because no significant flux or
spectral changes were observed during the pre- and post-flare epochs A ($0-252$\,day) and C ($294-374$\,day)
(see Figure\,4), the average fluxes and photon indices
for these time periods are reported. Note that the average fluxes of
these pre- and post-flare epochs are comparable, whilst the
corresponding photon indices differ significantly by $\Delta
\Gamma_{\gamma} \simeq 0.2$. Moreover, the observed spectral evolution
during one year of the {\it Fermi}-LAT exposure reveals a
hysteresis-like character, more clearly seen for the flaring period
(epoch B, with a time bin of 2 weeks), followed by a gradual flattening
in the subsequent decay phase. Further insights into the spectral
evolution of NGC\,1275 within the {\it Fermi}-LAT photon energy range
are provided by Figure\,6, which shows the two SEDs for pre- and
post-flare epochs A and C. Here the dotted line corresponds to the
best ``power-law with an exponential cutoff'' fit function determined
from an average $\gamma$-ray spectrum, as given in
Figure\,3. Interestingly, the difference between the two SEDs consists 
of an excess at photon energies $\varepsilon_{\gamma} \simeq 1-2$\,GeV,
with the low-energy $\gamma$-ray flux remaining essentially unchanged between
the two epochs. This implies that (1) the $\gamma$-ray variability in
NGC\,1275 (and possibly other radio galaxies) may be restricted to
$\geq$\,GeV photon energies, and that (2) the position of the peak in
the high-energy spectral component (in the $\nu F_{\nu}$ representation)
may change substantially even within the same object with no
accompanying significant flux changes. Note in this context that both 
the VERITAS and MAGIC non-detections were obtained
during the pre-flare epoch A ($164-206$\,day; see Acciari et
al. 2009b; Aleksi\'c et al. 2010). Hence, the emerging
conclusion is that \emph{it is not the total flux above $100$\,MeV
which should play a major role in triggering TeV observations of
steep-spectrum {\it Fermi}-LAT sources, but instead it is the flux and photon
index determined at higher photon energies ($\geq$\,GeV)}.

On the theoretical side, this conclusion could be possibly justified by
noting that after a new episode of injection of freshly accelerated
electrons into the emission zone (e.g., a downstream region of a
shock), higher energy electrons may lag 
behind the lower-energy electrons. In such a case, as
discussed previously in the context of blazar modeling (e.g., Kirk et
al. 1998; Sato et al. 2008), an ``anti-clockwise'' hysteresis in the
flux versus photon index plane may arise, similar to what we observed
in NGC\,1275 at $\gamma$-ray photon energies.

Here, we comment on the question if NGC\,1275 -- being a
representative example of a low-power radio galaxy
-- may be considered as a misaligned
blazar, most likely of the LBL type. Assuming a homogeneous
jet model, one should expect its jet Doppler factor $\delta =
\Gamma_{\rm j}^{-1} \, (1-\beta_{\rm j} \, \cos \theta)^{-1} =
\Gamma_{\rm j}^{-1} \, (1-\sqrt{1-\Gamma_{\rm j}^{-2}} \, \cos \theta)^{-1} \simeq
1-2$, for the typically expected jet viewing angle of $\theta \simeq
20^{\circ}-30^{\circ}$ and jet bulk Lorentz factors $\Gamma_{\rm j} \simeq
10$. Indeed, modeling of the broad-band emission of LBLs
(beamed counterparts of radio galaxies such as NGC\,1275 by assumption)
requires $\delta \sim \Gamma_{\rm j} \sim 10$. Thus, if the only 
difference between radio galaxies and blazars is due to the viewing
angle,  this should manifest in 
(i) different observed positions of the spectral peaks in
the $\nu F_\nu$ representation ($\nu \propto \delta$), (ii) different
observed variability patterns ($t_{var} \propto \delta^{-1}$, assuming
the emission region is a moving source), and finally in (iii) different
observed luminosities ($L_{obs} \propto \delta^4$ for a moving blob
case, or $\propto \delta^3/\Gamma_{\rm j}$ for the steady jet (see Sikora et
al. 1997).

Indeed, both the low- and high-energy
$\nu F_\nu$ peaks of NGC\,1275 are located at
substantially lower frequencies than those of typical LBLs. 
Taking the difference of beaming factors into account (i.e., 
$\delta_{\theta = 0}/\delta_{\theta = 20^{\circ}} \sim 10$), the
broad-band SED of the NGC\,1275 would be similar to the
SEDs of blazars such as BL Lacertae or 3C\,66A. Note however 
that the position of
the high-energy spectral peak may change substantially in a single
object even for comparable flux levels, at least in NGC\,1275, so
the diagnostics related to the location of the spectral peaks may not be
very conclusive. 

Variability as short 
as day timescales is often observed in LBLs. For example, during the historical
flare of BL Lacertae in 1997, correlated $\gamma$-ray and optical
flares were observed, with the $\gamma$-ray flux increasing by a factor of
$2.5$ within a day (Bloom et al. 1997). Similarly, daily variability has
been discovered in both {\it Fermi}-LAT and VERITAS observations of
3C\,66A (Reyes et al. 2009). Hence, in the simple
unification scheme outlined above, we should expect NGC\,1275 to vary in
$\gamma$-rays on timescales of several days. This, however, is difficult
to test even with the excellent sensitivity of the {\it Fermi}-LAT
instrument, due to the limited photon statistics for weekly time
bins. The analysis presented in this paper gives instead only a robust
upper limit $t_{\rm var} \le$ a few months. However, note that 
day-timescale variability has been detected at TeV photon energy range
for the M\,87 radio galaxy (Acciari et al. 2009c). Thus, more frequent
monitoring of NGC\,1275 by ground-based Cherenkov Telescopes would be 
valuable in this respect.

The observed photon flux of
 $F_{\rm \varepsilon_{\gamma}>100 MeV} = (2.19 \pm 0.13) \times
 10^{-7}$\,ph\,cm$^{-2}$\,s$^{-1}$ 
implies an \emph{observed} (isotropic) $\gamma$-ray
 luminosity of NGC\,1275, $L_{\gamma} \simeq 4 \pi d_{\rm L}^2 \,
 (\Gamma_{\gamma}-1) \, F_{\varepsilon_{\gamma}>\varepsilon_0} \,
 \int_{\varepsilon_0}^{\varepsilon_{\rm c}}
 (\varepsilon/\varepsilon_0)^{1-\Gamma_{\gamma}} \, d \varepsilon \sim
 10^{44}$\,erg\,s$^{-1}$, for $\Gamma_{\gamma} \simeq
 2.0-2.2$ and $\varepsilon_{\rm c} \simeq 42$\,GeV. This  is already 
comparable to the typical observed $\gamma$-ray luminosities
 of LBLs, which range between $L_{\gamma} \sim 10^{44} - 10^{46}$\,erg\,s$^{-1}$.
 Note in particular that in the framework of
 the simple unification scheme discussed above, the beamed analog of
 NGC\,1275 would then be characterized by the observed $\gamma$-ray
 luminosity larger by a factor of $[\delta_{\theta = 0}/\delta_{\theta =
 20^{\circ}}]^4 \sim 3 \times 10^4$ (or at least $[\delta_{\theta =
 0}/\delta_{\theta = 20^{\circ}}]^3 \sim 3 \times 10^3$) than this,
 i.e. $L_{\gamma} > 10^{47}$\,erg\,s$^{-1}$. Such luminosities are not
 expected for LBL-type blazars (see Abdo et al. 2010a). On the
 other hand, the observed $\gamma$-ray luminosity of NGC\,1275 is not
 energetically problematic, since the total \emph{emitted} $\gamma$-ray
 power in this source seems rather moderate \emph{as long as the
 emitting plasma moves with highly relativistic bulk velocities}, namely
 $L_{\rm \gamma,\,em} \simeq (\Omega_{\rm j}/4 \pi) \, L_{\gamma} \simeq
 L_{\gamma} / 4 \Gamma_{\rm j}^2 < 10^{42}$\,erg\,s$^{-1}$ for $\Gamma_{\rm j} \sim
 10$, where $\Omega_{\rm j} \simeq \pi \theta_{\rm j}^2$ is the solid angle defined by
 the jet opening angle $\theta_{\rm j}$, for which we assumed $\theta_{\rm j} \sim
 1/\Gamma_{\rm j}$. Such a relatively small emitted power would constitute less
 than $1\%$ of the total kinetic power of the NGC\,1275 jet, estimated
 by Dunn \& Fabian (2004) to be roughly $L_{\rm j} \sim (0.3-1.3) \times
 10^{44}$\,erg\,s$^{-1}$. Yet the problem of an unexpectedly large observed
 $\gamma$-ray luminosity of the beamed analog of NGC\,1275 remains, and
 poses a serious challenge to the simplest version of the AGN
 unification scheme. In this context, a viable explanation for
 this problem would be to postulate that the high-energy emission
 observed from ``misaligned'' blazars such as NGC\,1275 (or M\,87) is
 dominated not by a jet ``spine'' 
characterized by large bulk Lorentz factors
 ($\Gamma_{\rm j} \sim 10$ as is the case in bona-fide blazars), but by the
 slower jet boundary layers ($\Gamma_{\rm j} \sim$ few) as discussed by
 several authors (e.g., Celotti et al. 2001, Stawarz \& Ostrowski 2002, 
Ghisellini et al. 2005).  Alternatively, one may propose that the 
$\gamma$-ray emission observed from radio galaxies is not produced 
within the ``proper'' blazar  emission zone, but at larger distances 
from the active center characterized by slower bulk velocities 
(say, $\Gamma_{\rm j} \simeq$ few) of the emitting plasma.
Yet another possibility may be that the inner jets in NGC\,1275 
(and also in similar objects) are in general intrinsically 
less relativistic than the 
ones in bona-fide blazars; this would be consistent with the 
conclusions of Lister \& Marscher (1997), who argue that radio-loud AGN 
with nuclear jets characterized by $\Gamma_{\rm j} \geq 10$ must be 
rather rare among the general population. Whichever scenario is correct,
 the {\it Fermi} results seem to indicate that low-power radio galaxies 
are most likely not simple off-axis analogs of BL Lac objects in terms 
of their $\gamma$-ray properties.

\acknowledgments 

The \textit{Fermi}-LAT Collaboration acknowledges generous ongoing 
support from a number of agencies and institutes that have supported 
both the development and the operation of the LAT as well as scientific 
data analysis. These include the National Aeronautics and Space 
Administration and the Department of Energy in the United States, the 
Commissariat \`a l'Energie Atomique and the Centre National de la 
Recherche Scientifique / Institut National de Physique Nucl\'eaire et de 
Physique des Particules in France, the Agenzia Spaziale Italiana and the 
Istituto Nazionale di Fisica Nucleare in Italy, the Ministry of 
Education, Culture, Sports, Science and Technology (MEXT), High Energy 
Accelerator Research Organization (KEK) and Japan Aerospace Exploration 
Agency (JAXA) in Japan, and the K.~A.~Wallenberg Foundation, the Swedish 
Research Council and the Swedish National Space Board in Sweden.

Additional support for science analysis during the operations phase is 
gratefully acknowledged from the Istituto Nazionale di Astrofisica in 
Italy and the Centre National d'\'Etudes Spatiales in France.

We acknowledge S.~Digel and  J.~Finke for their helpful comments 
to improve the manuscript.  
\L.~S.\ is grateful for the support from the Polish MNiSW through the grant 
N-N203-380336.

\begin{figure}
\begin{center}
\includegraphics[angle=0,scale=0.8]{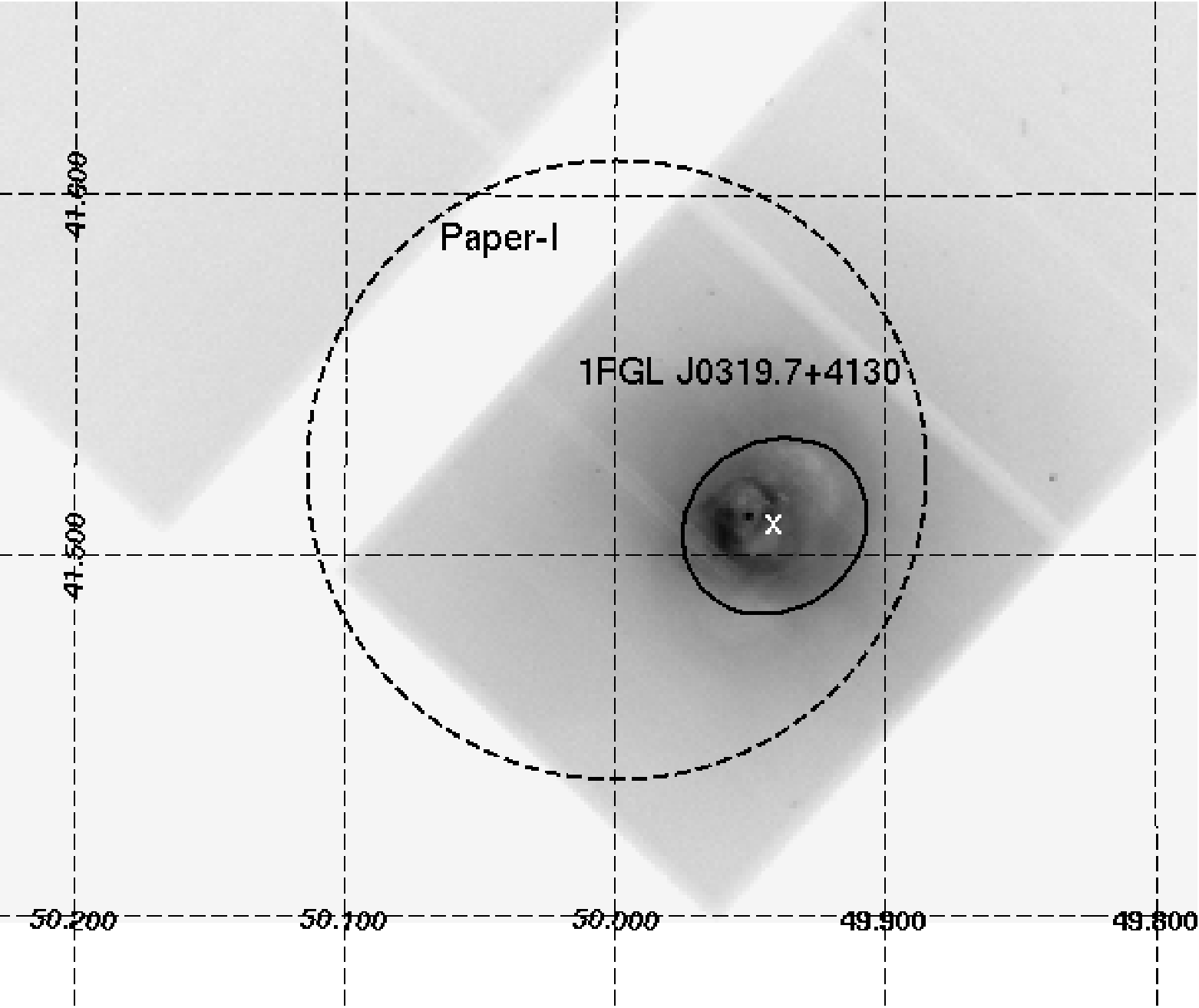}
\caption{The 95 $\%$ LAT $\gamma$-ray localization error circles
placed on the $Chandra$ ACIS-S image constructed between 0.4 and 8.0
 keV. The positional center of the $\gamma$-ray emission, marked as the
 white ``X'', is only 0.46 arcmin
 from the position of the NGC~1275 nucleus, with 95$\%$ radii of
$r_{\rm 95}$ = 5.2 arcmin for Paper-I and 1.5 arcmin (more accurately,
1.56 $\times$ 1.38 arcmin; Abdo et al. 2010b) for 11 months, respectively.}\label{fig:X_map}
\end{center}
\end{figure}

\begin{figure}
\begin{center}
\includegraphics[angle=0,scale=0.8]{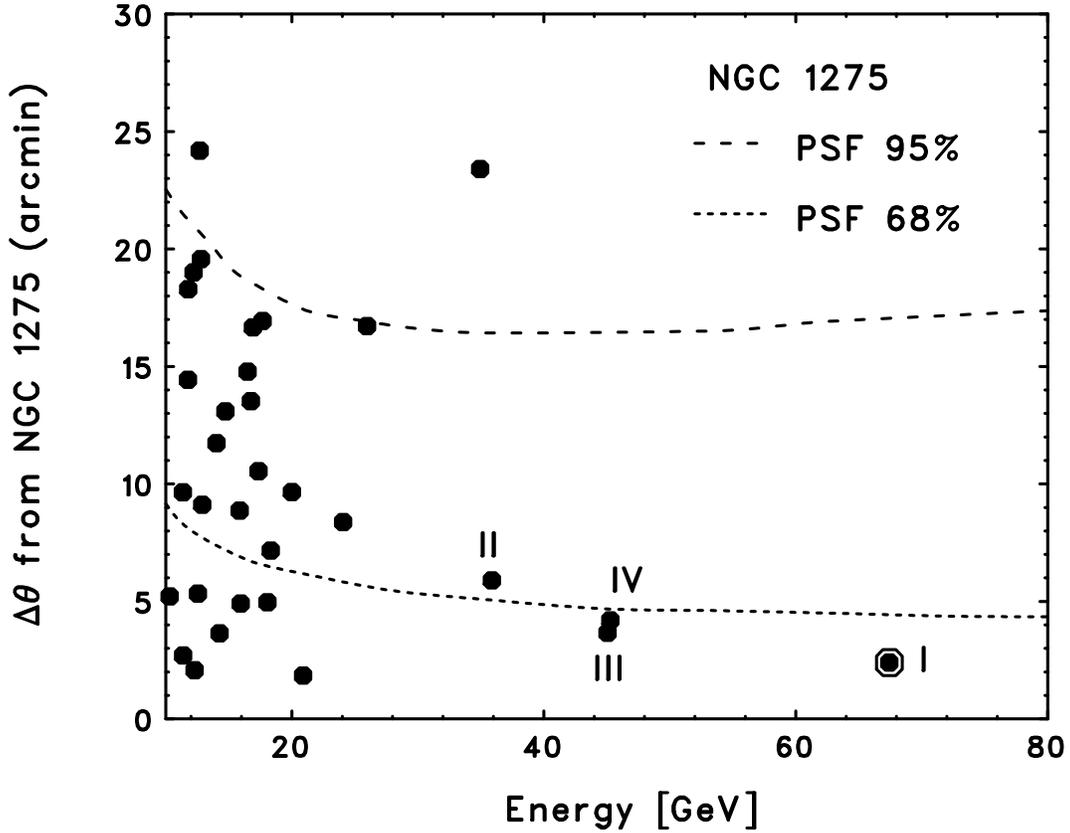}
\caption{Angular separation of high energy ($\varepsilon_{\gamma} > 10$\,GeV) photons
from the nucleus of NGC\,1275 as a function of photon energy. The $68\%$
and $95\%$ angular resolutions of {\it Fermi}-LAT are shown as dotted
and dashed lines, respectively. These PSF profiles are derived for 
{P6\_V3\_Diffuse} and have been averaged over the acceptance of the LAT.
The highest energy photon detected during
the one-year-all-sky survey is $\varepsilon_{\gamma} = 67.4$\,GeV ({\it
double circle} denoted as ``I''), whose angular separation from
NGC\,1275 is only $2.4$\,arcmin. Arrival times of the four highest energy
photons (``I-IV''), are indicated in
Figure\,4.}\label{fig:angle_sep}
\end{center}
\end{figure}

\begin{figure}
\begin{center}
\includegraphics[angle=0,scale=0.8]{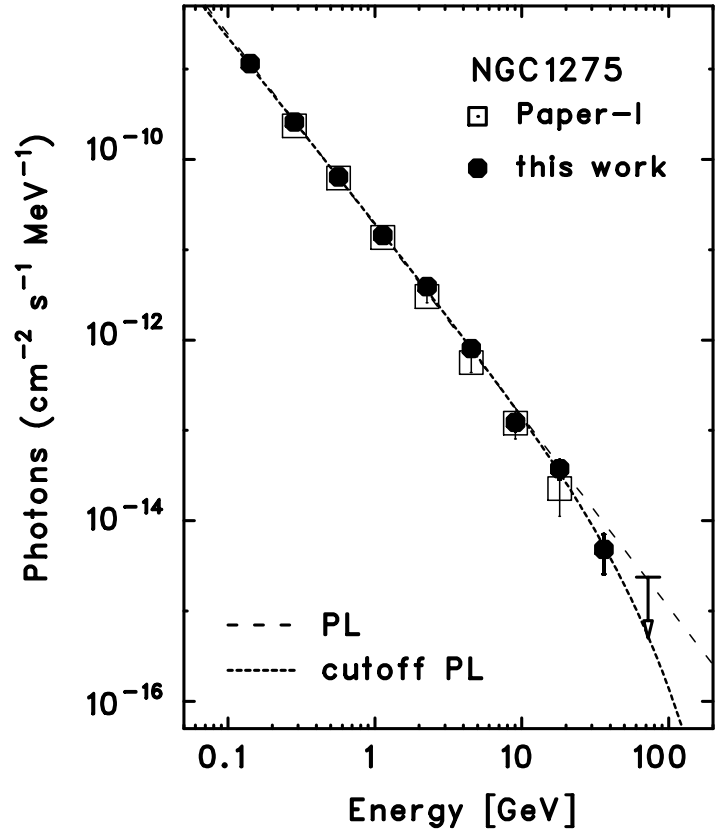}
\caption{The average LAT $>$100 MeV spectrum of NGC\,1275 derived 
from one year accumulation of data ({\it filled circles}; this work), as
compared with that determined from the initial 4-month dataset ({\it open squares};
Paper I). The dashed line shows the power-law function for the one-year data
determined from the $\textsc{gtlike}$. The dotted line represents the
best-fit cutoff power-law with $\varepsilon_{\rm c} = 42.2$\,GeV as
described in the
text.}\label{fig:LAT_spec}
\end{center}
\end{figure}

\begin{figure}
\begin{center}
\includegraphics[angle=0,scale=0.8]{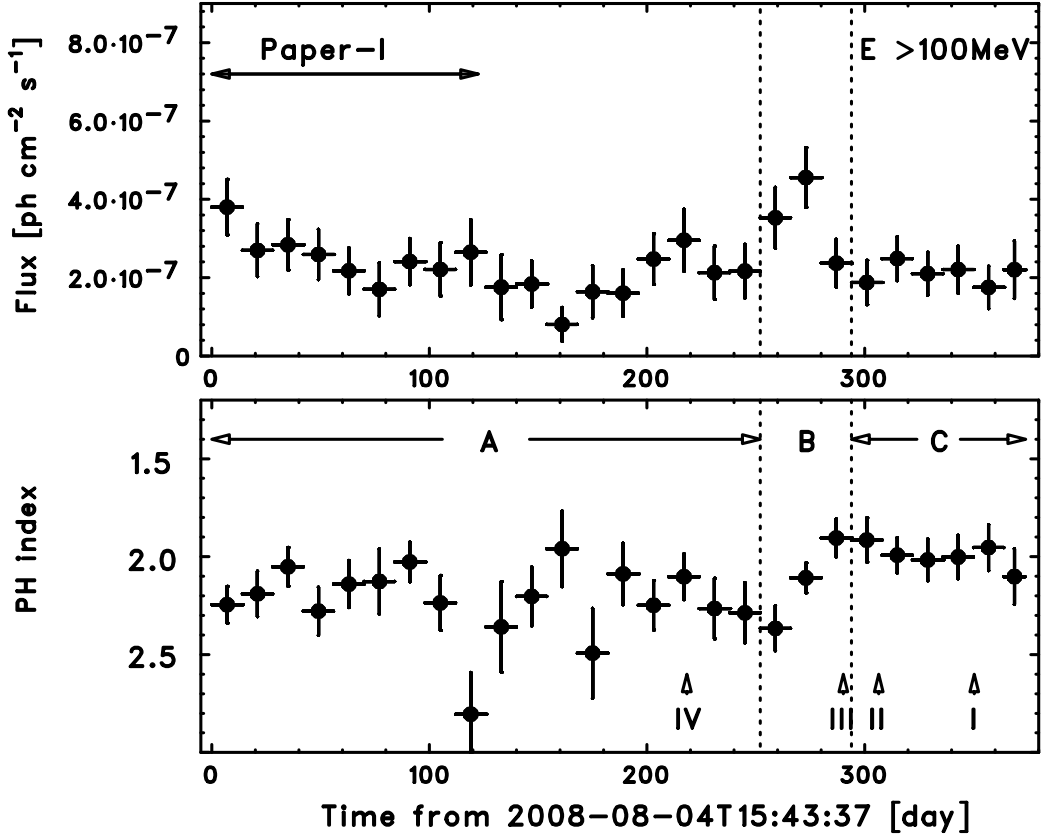}
\caption{Temporal variation of $\gamma$-ray flux and spectral index 
over the period August 2008 -- August 2009. The time (in days) is 
measured from the start of the {\it Fermi} observation, i.e., 2008 
August 4, 15:43:37 UT. {\bf Upper panel:} changes in the 
$\varepsilon_{\gamma} > 100$\,MeV flux. {\bf Lower panel:} changes 
in the power-law photon index. We have divided the analyzed time 
window into epochs A (before the flare), B (during the flare), and C 
(after the flare). The arrival times of the highest energy photons 
I-IV (Figure\,2) are indicated as arrows. Background diffuse 
emission (both Galactic and extra-galactic) is fixed at the best-fit 
parameters determined from an average spectral fit as given in the 
text, and only statistical errors are shown.}\label{fig:LAT_LC}
\end{center}
\end{figure}

\begin{figure}
\begin{center}
\includegraphics[angle=0,scale=0.8]{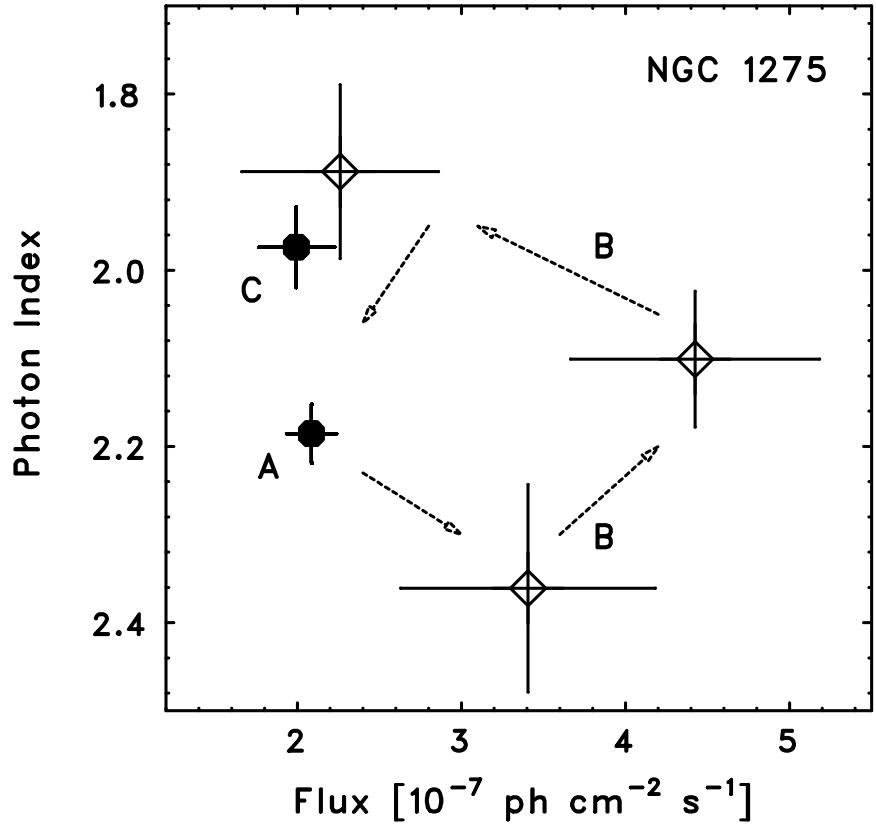}
\caption{Spectral
evolution of NGC\,1275 in the flux--photon index representation. Labels
``A'' and ``C'' denote the average fluxes and photon indices determined
for the period before ($0-252$\,day) and after the flare
($294-374$\,day), respectively (see Figure\,4). Label ``B'' denotes the
spectral evolution during the flare with a time bin of 2 weeks. Note
that the photon index changes significantly before and after the flare, 
while the flux levels are comparable.}
\end{center}
\end{figure}

\begin{figure}
\begin{center}
\includegraphics[angle=0,scale=0.8]{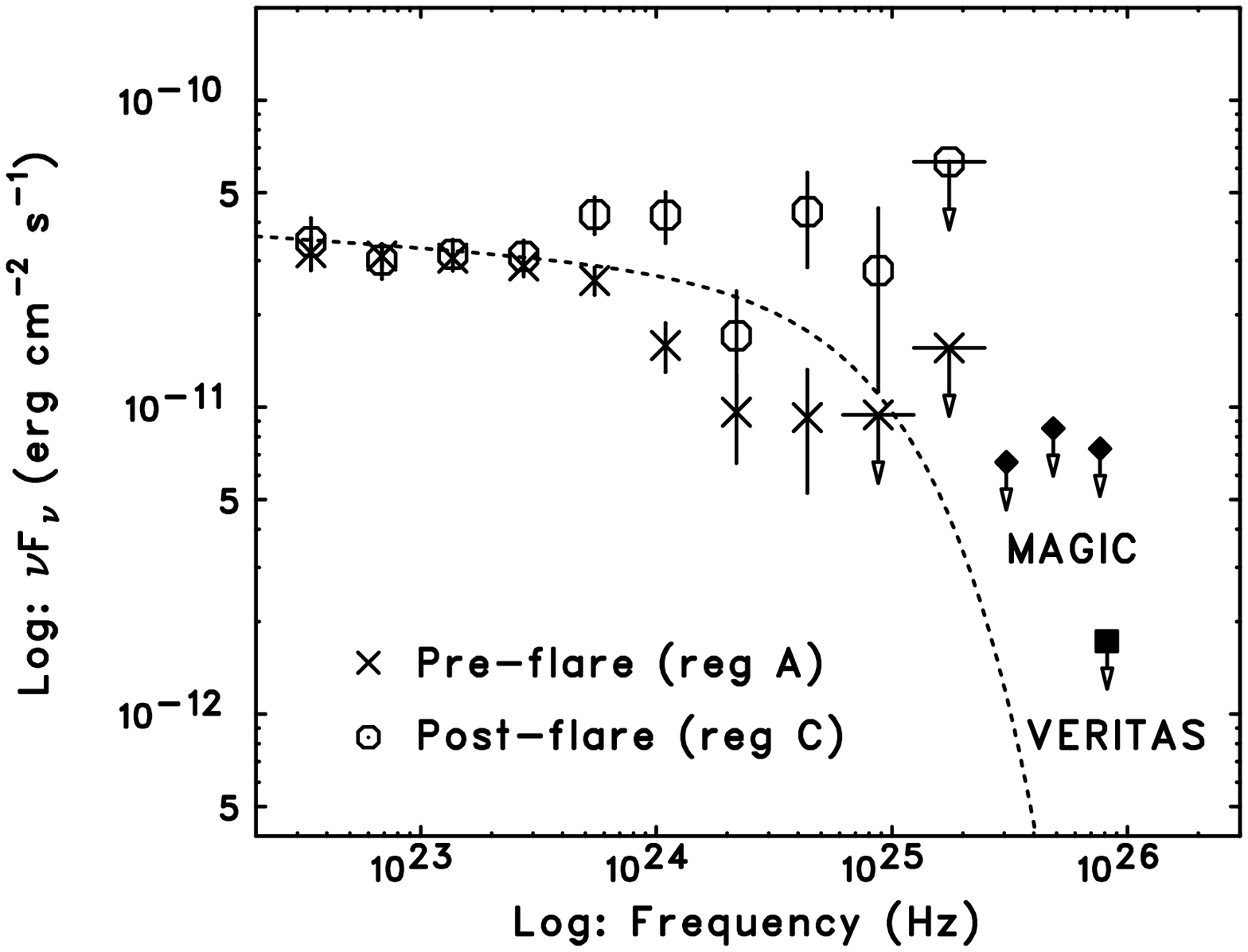}
\caption{The $\gamma$-ray spectral energy distribution of NGC\,1275 before (crosses)
and after the flare (open circles), corresponding to epochs A and C in
Figure\,4, respectively. Note that the excess high-energy $\gamma$-ray
emission only appears above $\varepsilon_{\gamma} \simeq 1-2$\,GeV. The
flux upper limits in the TeV range resulting from the VERITAS and MAGIC
observations correspond to the pre-flare epoch A. The dotted line
represents the best-fit cutoff power-law fit as given in Figure\,3.}\label{fig:LAT_spec2}
\end{center}
\end{figure}


\begin{thebibliography}{}
\bibitem[]{} Abdo, A., et al. ({\it Fermi}-LAT Collaboration) 2009a, \apj, 699, 31 (Paper-I) 
\bibitem[]{} Abdo, A., et al. ({\it Fermi}-LAT Collaboration) 2009b, \apj, 700, 597 (LAT Bright AGN Sample)
\bibitem[]{} Abdo, A., et al. ({\it Fermi}-LAT Collaboration) 2009c, \apjs, 183, 46 (LAT Bright Source List)
\bibitem[]{} Abdo, A., et al. ({\it Fermi}-LAT Collaboration) 2009d, \apj, 707, 55 (M87)
\bibitem[]{} Abdo, A., et al. ({\it Fermi}-LAT Collaboration) 2010a, \apj, submitted (arXiv:1002.0150v1)
\bibitem[]{} Abdo, A., et al. ({\it Fermi}-LAT Collaboration) 2010b, \apjs, submitted (arXiv:1002.2280v1)
\bibitem[]{} Acciari, V.~A., et al. 2008, \apjl, 684, L73
\bibitem[]{} Acciari, V.~A., et al. 2009a, \apjl, 693, L104
\bibitem[]{} Acciari, V.~A., et al. 2009b, \apjl, 706, L275
\bibitem[]{} Acciari, V.~A., et al. 2009c, Science, 325, 444
\bibitem[]{} Acciari, V.~A., et al. 2009d, \apj, 707, 612
\bibitem[]{} Ajello, M., et al. 2009, ApJ, 690, 367
\bibitem[]{} Albert, J., et al. 2007, \apjl, 666, L17
\bibitem[]{} Aleksi\'c, J., et al. 2010, \apj, 710, 634
\bibitem[]{} Anderhub, H., et al. 2009, \apjl, 704, L129
\bibitem[]{} Asada, K., et al, 2006, PASJ, 58, 261
\bibitem[]{} Atwood, W.~B., et al. ({\it Fermi}-LAT Collaboration) 2009, \apj, 697, 1071
\bibitem[]{} Balmaverde, B., Capetti, A., \& Grandi, P. 2006, \aap, 451, 35
\bibitem[]{} Bloom. S.~D., et al. 1997, \apj, 490, L145
\bibitem[]{} B\"{o}hringer, H., et al. 1993, \mnras, 264, L25
\bibitem[]{} Celotti, A., Ghisellini, G., \& Chiaberge, M. 2001, \mnras, 321, L1
\bibitem[]{} Dunn, R.~J.~H., \& Fabian, A.~C. 2004, \mnras, 355, 862
\bibitem[]{} Fabian, A.~C., et al. 2003, \mnras, 344, L43
\bibitem[]{} Fabian, A.~C., et al. 2006, \mnras, 366, 417
\bibitem[]{} Ghisellini, G., Celotti, A., Fossati, G., Maraschi, L., \& Comastri, A. 1998, \mnras, 301, 451
\bibitem[]{} Ghisellini, G., Tavecchio, F., \& Chiaberge, M. 2005, \aap, 432, 401
\bibitem[]{} Kirk, J.~G., \& Rieger, F.~M., \& Mastichiadis, A. 1998,
	  \aap, 333, 452
\bibitem[]{} Lister, M. L., \& Marscher, A. P. 1997, \apj, 476, 572
\bibitem[]{} Mayer-Hasselwander, H.~A., et al. 1982, \aap, 105, 164
\bibitem[]{} Owen, F.~N., Eilek, J.~A., \& Kassim, N.~E. 2000, \apj, 543, 611
\bibitem[]{} Reimer, O., Pohl, M.,Sreekumar, P., \& Mattox, J.~R. 2003, \apj, 588, 155
\bibitem[]{} Reyes, L. C., et al. 2009, in Proceedings of the 31st ICRC (astro-ph/0907.5175)
\bibitem[]{} Sato, R., et al. 2008, \apj, 680, L9 
\bibitem[]{} Sikora, M., Madejski, G., Moderski, R., \& Poutanen, J. 1997, \apj, 484, 108
\bibitem[]{} Stawarz, {\L}., \& Ostrowski, M. 2002, \apj, 578, 763
\bibitem[]{} Strong, A.~W., et al. 1982, \aap, 115, 404
\bibitem[]{} Taylor, G.~B., \& Vermeulen, R.~C. 1996, \apjl, 457, L69
\bibitem[]{} Vermeulen, R.~C., Readhead, A.~C.~S., \& Backer, D.~C. 1994, \apjl, 430, L41
\bibitem[]{} Walker, R.~C., et al. 2000, \apj, 530, 233
\bibitem[]{} Weekes, T.~C., et al. 2002, Astropart. Phys., 17, 22
\end{thebibliography}
\end{document}